\documentclass[figures,cite,nobm]{epl}
\usepackage{epsf}
\newcommand {\dist}{{\rm dist}}

\title{Emergence of memory }
\author{Konstantin Klemm and Preben Alstr\o m}
\institute{Niels Bohr Institute, Blegdamsvej 17, DK-2100 Copenhagen \O ,
Denmark}
\begin{document}
\maketitle
\begin{abstract}
We propose a new self-organizing mechanism behind the emergence of memory
in which temporal sequences of stimuli are transformed into spatial
activity patterns. In particular, the memory emerges despite the absence
of temporal correlations in the stimuli. This suggests that neural systems
may prepare a spatial structure for processing information before the information itself is available. A simple model illustrating the mechanism is presented based on three principles: (1) Competition between neural units, (2) Hebbian plasticity, and (3) recurrent connections.
\end{abstract}

Memory is believed to be a universal feature of the nervous system
\cite{Thorpe63} and exciting results improving our understanding of
molecular as well as organizational mechanisms underlying memory have been
obtained in recent decades \cite{Kandel00}. On the organizational level
significant work has been devoted to the study of ``brain maps''
underlying the ability to recognize patterns or features from a given
sensory input \cite{HertzKroghPalmer,Kohonen01}. Many intriguing
suggestions have been given as to how a memory emerges that is able to
extract and recall features from a spatial pattern of neural activity
\cite{Hampson99,Kohonen99}.

In this Letter, we focus on the mechanism behind self-organization
from a {\em temporal} sequence of activity. 
Time is important in many cognitive tasks, e.g.\
vision, speech, signal processing and motor control.
The crucial point is how to represent time, and methods
often involve time delays in one form or another
\cite{Unnikrishnan91,Haykins99}. How does a structured memory emerge
that can cope with temporal sequences of activity? 
For example, the information
we receive through a temporal sequence of input must at least to some
extent be memorized spatially in the neuronal activity pattern.
Here we present a simple conceptual
model for the time to space transformation, from which a memory emerges.

The fundamental assumptions of the model presented here are the following:
(1) {\em Competition} between neural units; excited neural units have an
inhibiting effect on other units. In the limit of strong inhibition this is
winner-take-all \cite{WTA} where only the region of units with the
strongest excitation remains active, suppressing all surrounding units.
(2) {\em Hebbian Plasticity} is an abstract formulation of long term
potentiation depending on pre- and postsynaptic activity: If activity of
unit A is followed by activity of unit B the connection from A to B is
strengthened \cite{Bliss93,Hebb49}. (3) {\em Recurrent} connectivity opens
up the possibility for ongoing information processing in the network by
internal feedback.

Features (1) and (2) mentioned above are employed by the {\em
self-organizing map} model formulated by Kohonen \cite{Kohonen01}.
Recently it has been argued \cite{Kohonen99} that the self-organizing map
is a biologically plausible large-scale model of cortical information
processing. However, the self-organizing map has a purely unidirectional
information flow without internal dynamics. We know of few attempts to
explicitly integrate memory of past stimuli into the self-organizing map
\cite{Kangas90,Somervuo99,Koskela98,Liu99,Euliano98}. These approaches
have been shown to work well on specific tasks.

The scope of the current paper is to investigate generally,
i.e.~task-independent, the formation of an internal dynamics that can
lead to formation of memory. In our approach, memory is not designed but
emerges as a result of the self-organized dynamics of the neural system.

Consider $M$ neural units arranged as a one-dimensional lattice with
periodic boundary conditions (a ring). The model describes the
time-discrete evolution of the real-valued activities
$y_0(t),\dots,y_{M-1}(t)$ of the units.
At a given time step $t$ each unit $i$ receives a recurrent excitation
$h_i^{\rm rec}(t)=\sum_j w_{ij} y_j(t-1)$ through connections $w_{ij}$.
Additionally there is an $S$-dimensional input ${\bf x} =
\left(x_1(t),\dots,x_S(t)\right)$ to the system causing an external
excitation $h_i^{\rm ext}(t)=\sum_j v_{ij} x_j(t)$ through connections
$v_{ij}$.
The total excitation is $h_i(t)=h_i^{\rm rec}(t)+h_i^{\rm ext}(t)$. Next,
we define the centre of activity $i^\ast$ as the unit with the largest
total excitation: $i^\ast(t) = \arg \max_i h_i(t)$. The updated unit
activities form a Gaussian profile around the centre of activity (we
suppress $t$ in the notation here)
\begin{equation}
y_i = c \exp \left(-\frac{\dist^2(i,i^*)} {2\sigma^2}\right)~,
\end{equation}
where \dist(i, i*)
denotes the distance between units $i$ and $i^\ast$ in lattice points. The
model parameter $\sigma$ is a measure of the width of the neural activity
field. The constant value $c>0$ is chosen such that the normalization
$\sum_i (y_i)^2 = 1$ holds. 
Finally, all connections are updated according to a Hebb-rule with a
saturation term. Each recurrent connection $w_{ij}$ is changed by
\begin{equation}
\Delta w_{ij} = \eta y_i(t) \left(y_j(t-1) -
w_{ij}\right)~,
\end{equation}
where $\eta>0$ is a constant learning rate.
Correspondingly, the increment for the input connections is
\begin{equation}
\Delta v_{ij} = \eta y_i(t)\left(x_j(t)- v_{ij}\right)~.
\end{equation}
This completes one time step of the dynamics.
The learning rate has a value $\eta = 0.2$ in all the simulations
presented in the following. The length scale is taken to be $\sigma=1.0$. The
connections $w_{ij}$ and $v_{ij}$ are initialized with random values in
the interval $[0; 0.001]$.

The memorization ability of the network is the degree to which the state
of the network, given by $i^\ast$, depends on the past stimuli. A suitable
measure of statistical dependence between the two stochastic variables is
their mutual information \cite{Shannon63}. Given a discrete set $X$ of
possible stimuli, the mutual information between the current centre of
activity $i^\ast(t)$ and the stimulus ${\bf x}(t-\tau)$ presented $\tau$
time steps before reads
\begin{equation} \label{mutinf}
T_\tau = \sum_{i=1}^M \sum_{{\bf x^\prime}\in X}
p_\tau(i,{\bf x^\prime})
         \log_2 \frac {p_\tau(i,{\bf x^\prime})}
         {\Pr(i^\ast=i) \Pr({\bf x}={\bf x^\prime})}
\end{equation}
where $p_\tau(i,{\bf x^\prime})= \Pr(i^\ast(t)=i \wedge {\bf
x}(t-\tau)={\bf x^\prime})$ is the joint distribution of the centre of
activity and the past stimulus.
When estimating the joint probability distribution $p_\tau$ and its
marginals for a given network at a certain time, the dynamics is sampled
over 5000 time steps with $\eta=0$. Consequently these time steps are not
included in the learning time measured.

Let us now demonstrate the emergence of memory by simulations where the
network is presented with a random time series. The considered networks
have $M=64$ units and $S=2$ inputs. We present only two different
orthogonal stimuli, ${\bf x}=(2,0)$ and ${\bf x}=(0,2)\}$. We use ${\bf
0}$ and ${\bf 1}$ as shorthand for the two stimulus vectors.  At each time
step one of the vectors ${\bf 0}$ and ${\bf 1}$ is drawn randomly with
probability $p=0.5$.

\begin{figure}[ht]
\let\picnaturalsize=N
\def\picsize{100mm} 
\def\picfilename{fig1.eps}
\let\epsfloaded=Y
\centerline{\ifx\picnaturalsize N\epsfxsize \picsize\fi
\epsfbox{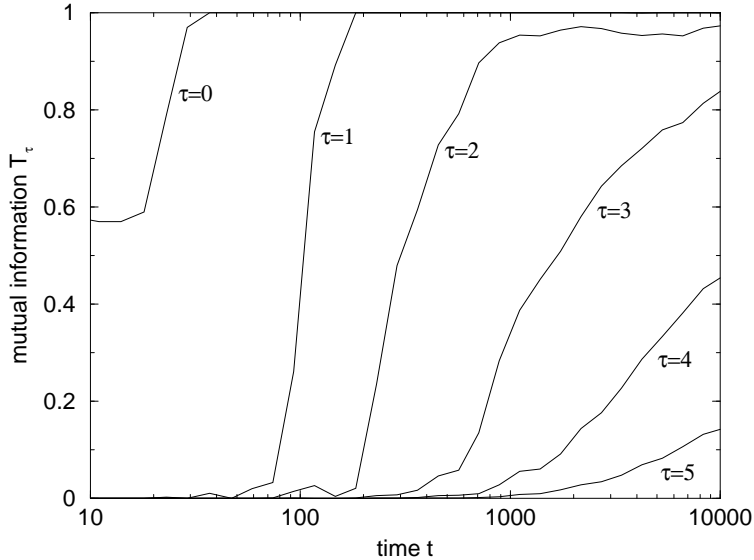}}
\caption{The time evolution of the mutual information $T_\tau$ indicates
the formation of memory. The stream of stimuli ${\bf x}(t)$ contains
1 bit of information per time step (two different stimuli presented with
equal probability). Thus $T_{\tau}=1$ means that the network perfectly
remembers the stimulus presented $\tau$ time steps before whereas $T_\tau=0$
means statistical independence between the stimulus and the network state.
The displayed
results were obtained as averages over 100 independent simulation runs
with networks of size $M=64$ units. \label{fig1}
}
\end{figure}
Fig.~\ref{fig1} shows the time evolution of the mutual information
$T_{\tau}$. Originally the state of the network depends only on the
current stimulus. This means $T_0>0$, but $T_\tau=0$ for all $\tau>0$.
After approximately 40 time steps the two stimuli are always discriminated
by different network states ($T_0=1$). Before step $t=200$ we observe the
emergence of memory: $T_1=1$ indicates full discrimination between stimuli
presented the previous time step. With further learning the memory length
expands to more time steps, hence $T_2>0$, $T_3>0$ and so on. The maximum
information content of the network is bounded by the number $M$ of
possible states (centres of activity). Thus the condition
$\sum_{\tau=0}^\infty T_\tau \leq\log_2 M = 6$ causes a saturation in the
formation of the memory.

\begin{figure}[ht]
\let\picnaturalsize=N
\def\picsize{100mm} 
\def\picfilename{./fig2.eps}
\let\epsfloaded=Y
\centerline{\ifx\picnaturalsize N\epsfxsize \picsize\fi
\epsfbox{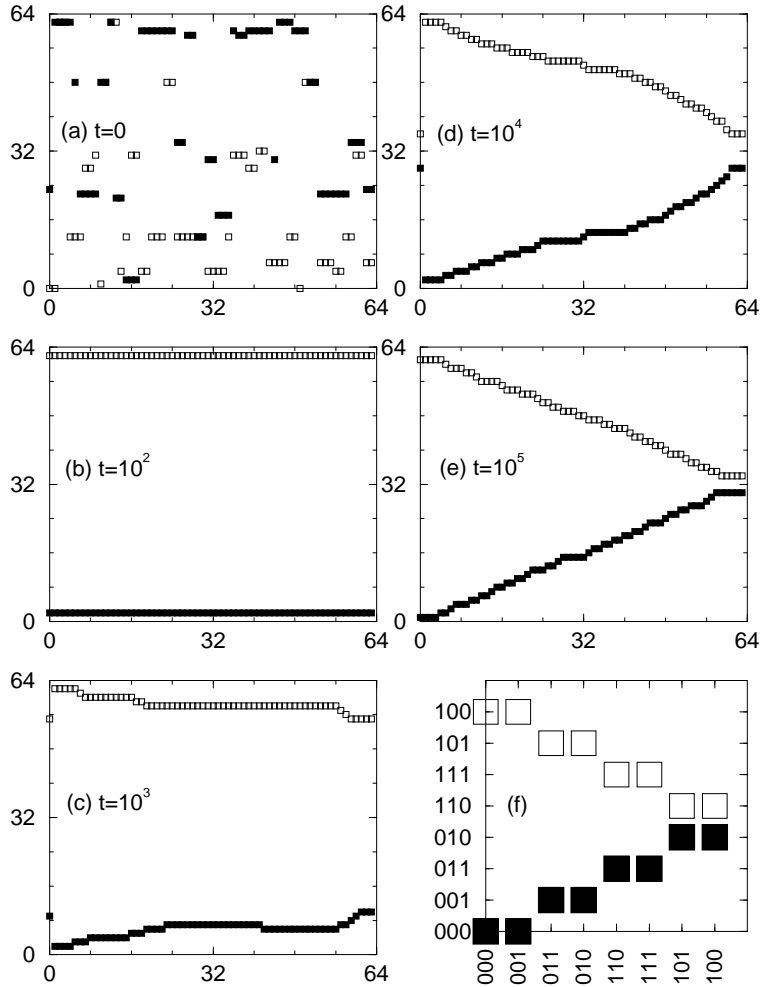}}
\caption{Spatial representation of memory. (a)-(e) The return-map of
the centre of activity $i^\ast$
after $0,\dots,10^5$ steps of learning. The diagrams show $i^\ast (t)$
as a function $i^\ast (t-1)$. The function has two branches (filled and 
unfilled squares) corresponding to the two different values the stimulus
${\bf x}(t)$ can assume. (f) Idealized return-map for a network with
$M=8$ units. Each unit represents a certain history of stimuli. The histories
of the units are given as bit strings on the axes.
\label{fig2}
}
\end{figure}
More insight can be gained by considering the geometrical structure of the
memory. In Fig.~\ref{fig2} we have plotted the evolution of the return-map
of the network dynamics for a typical simulation run configured as in the
previous section. The diagrams are to be interpreted as follows: The abscissa
is the centre of activity $i^\ast(t-1)$ in the previous time step. The
ordinate is the subsequent centre of activity $i^\ast(t)$. Depending on the
given stimulus $x(t)$ either the filled or the unfilled squares represent the
mapping $i^\ast(t-1)\rightarrow i^\ast(t)$. We observe that the two branches
of the return map tend to become straight lines with slopes 1/2 and -1/2,
respectively. Panel (f) of FIG.~\ref{fig2} shows an idealized version for
the case of $M=8$ units. The emerging return map $f$ can be interpreted as the
inverse of a tent map where the ambiguity of the two branches is resolved by
the given stimulus.
\begin{equation} \label{returnmap1}
i^\ast \mapsto f_{\bf x}(i^\ast) = \left\{
\begin{array}{ll}
\lfloor i^\ast/2 \rfloor, & {\rm if} \quad {\bf x}={\bf 0} \\
M-1-\lfloor i^\ast/2 \rfloor, & {\rm if} \quad {\bf x}={\bf 1} \\
\end{array} \right.
\end{equation}
By $\lfloor . \rfloor$ we denote the integer part of the argument. 
In order to understand how the stimuli are stored in the network state,
it is convenient to write the centre of activity as a binary number
$i^\ast = \sum_{k=0}^{L-1} 2^k i_k =: (i_{L-1},\dots,i_0)$, where
$L=\log_2 M$ denotes the number of bits used. Writing also the stimulus
${\bf x}$ as a binary value $x\in\{0,1\}$, the return map Eq.\
(\ref{returnmap1}) reads
\begin{equation}
f_x(i_{L-1},\dots,i_0)=
  (x,i_{L-1}\oplus x,i_{L-2}\oplus x,\dots,i_1\oplus x)
\end{equation}
The operation $\oplus$ is the exclusive-or ($a\oplus b=0$ if
$a=b$, otherwise $a\oplus b=1$). Thus the operation $f_0$ shifts all
bits of the argument to the right, discards the least significant bit
and inserts $x$ as the highest significant bit. $f_1$ additionally
inverts all bits of the argument. Applying $f_x$ iteratively $\tau
\ge L$ times we obtain
\begin{eqnarray}
i^\ast(t) & = &
f_{x(t)} \circ f_{x(t)} \circ \dots \circ f_{x(t-\tau+1)} (i^\ast(t-\tau))\\
          & = & 
\left(x(t),x(t-1)\oplus x(t),x(t-2)\oplus x(t-1)\oplus x(t),..., \bigoplus_{s=0}^{L-1} x(t-s)\right)
\end{eqnarray}
Thus at any time $t$ the values $x(t),x(t-1),\dots,x(t-L+1)$ of the
$L$ previous stimuli can be extracted from $i^\ast(t)$.
Note that, due to the non-linear superposition of the stimuli by the
exclusive-or, the memory effect in general cannot be observed when applying
purely linear measures. In particular, the linear correlation function
between $i^\ast(t)$ and $x(t-\tau)$ vanishes for $\tau>0$. However,
using the mutual information $T_\tau$ (Eq.\ (\ref{mutinf})) one detects
the memorization of past stimuli in $i^\ast(t)$. 
For more than two discrete stimuli the emergence
of memory is observed accordingly, forming a return map with more than
two branches.

We now consider the case of asymmetry in the presentation of stimuli. We
use the same two stimuli as in the preceeding sections. Unlike before, we
admit the probability $p$ of presenting stimulus ${\bf 1}$ to assume values
different from the symmetric case $p=0.5$. The amount of information
per time step in the stream of stimuli is then given by the Shannon function
$S(p)= p\log_2(p) + (1-p)\log_2(1-p)$.

\begin{figure}
\let\picnaturalsize=N
\def\picsize{100mm} 
\def\picfilename{./fig3.eps}
\let\epsfloaded=Y
\centerline{\ifx\picnaturalsize N\epsfxsize \picsize\fi
\epsfbox{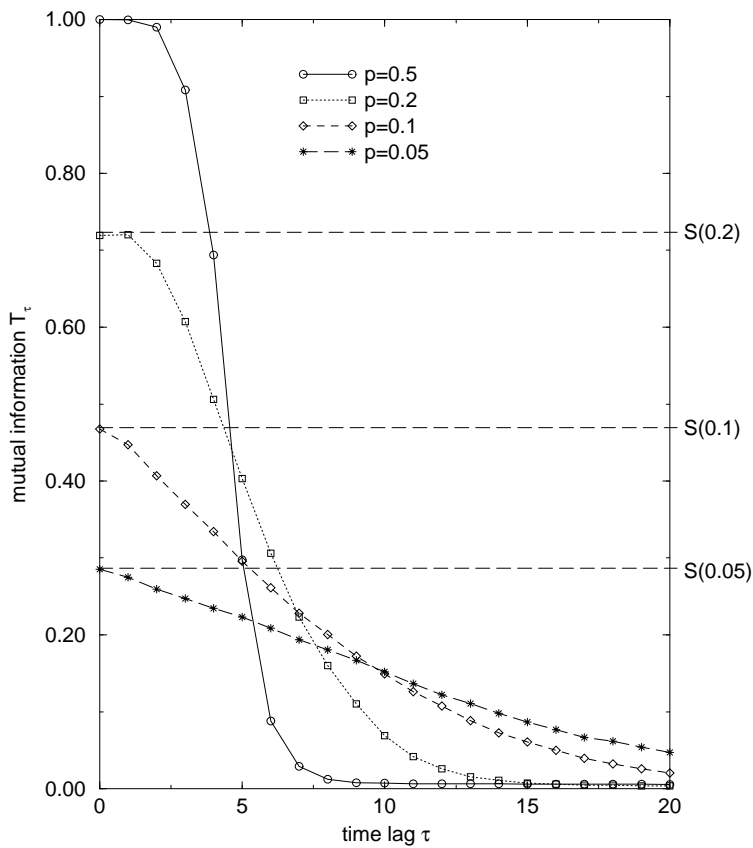}}
\caption{\label{fig3}Adaptation to asymmetry in the
occurence of the stimuli, as the probability $p$ of presenting the
stimulus ${\bf 1}$ deviates from 0.5.
The network adapts to the value of $p$ by varying the memory length. For
$p=0.5$ only a few time steps can be remembered with the available number
of units. Lowering $p<0.5$ decreases the information per time step in the
stream of stimuli. Then longer memories are possible. Each plotted value is
an average over 100 independent runs with networks of size $M$. The networks
had learnt for 100,000 time steps before mutual information was estimated.
}
\end{figure}

Figure \ref{fig3} shows the mutual information as a function of the time-lag
$\tau$ for different values of $p$. For small $\tau$ the mutual information
is close to $S(p)$ for all considered values of $p$. This means that in any
case the network almost perfectly memorizes a few preceeding time steps.
However, varying $p$ causes a redistribution of memory: As the parameter $p$
decreases, the decay of $T$ with growing $\tau$ becomes weaker: The smaller
$p$, the ``longer'' the memory. Thus the neural network
automatically adapts to the statistics of the stimuli.

\begin{figure}
\let\picnaturalsize=N
\def\picsize{100mm} 
\def\picfilename{./fig4.eps}
\let\epsfloaded=Y
\centerline{\ifx\picnaturalsize N\epsfxsize \picsize\fi
\epsfbox{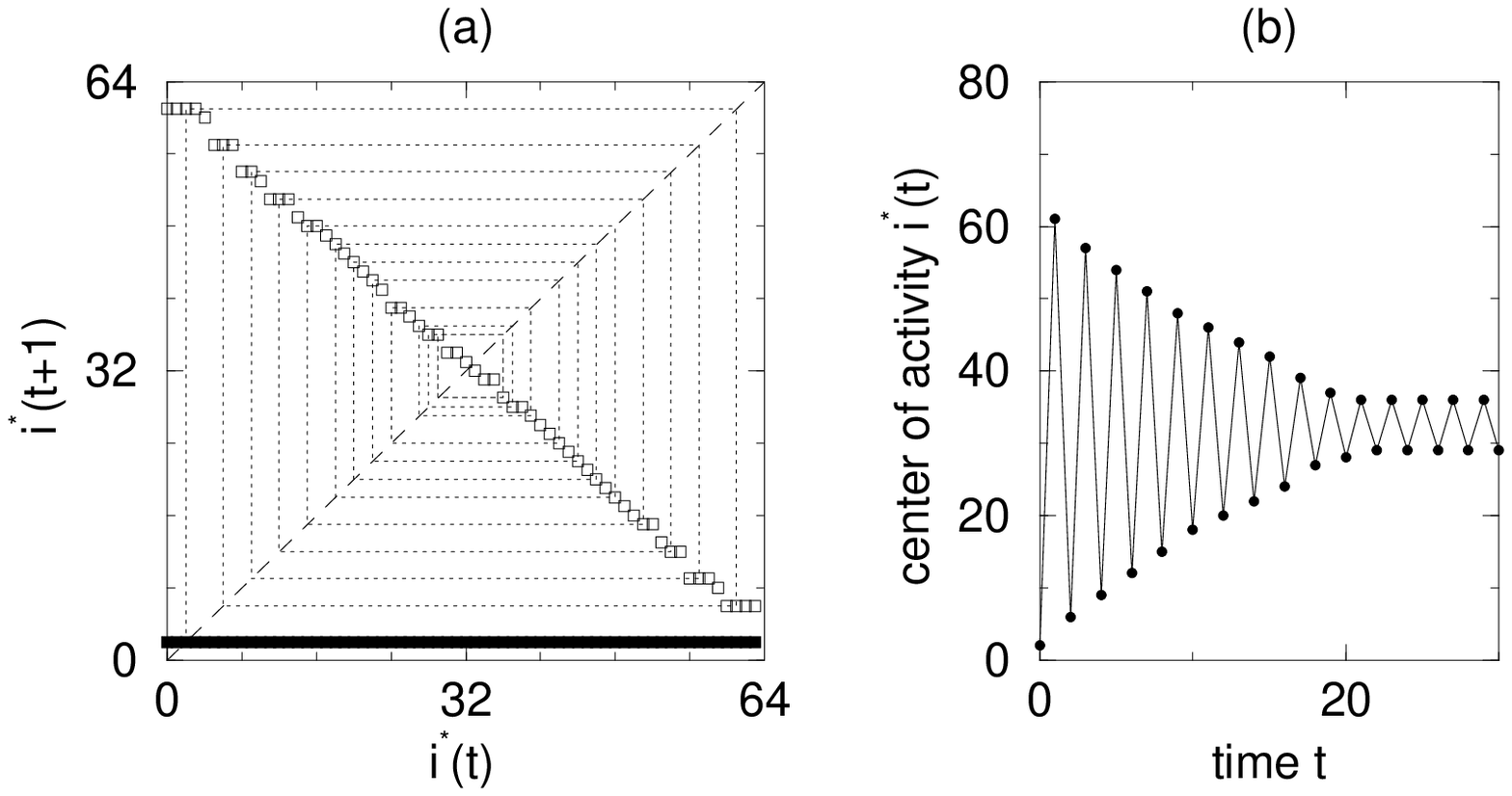}}
\caption{\label{fig4}
Typical simulation results for the case that one of the two stimuli occurs
very seldomly (here with probability $p=0.05$). (a) Return map of $i^\ast$,
after $10^6$ learning steps. When the seldom stimulus ${\bf x}(t)={\bf 1}$
occurs, $i^\ast(t)=2$ becomes the centre of activity (filled horizontal
branch).
Presentation of the other stimulus ${\bf x}={\bf 0}$ in the
subsequent time steps leads to the iteration dynamics indicated by the
dotted lines. (b) Corresponding time series of the centre of activity $i^\ast$.
20 time steps after the last presentation of stimulus ${\bf 1}$ the
dynamics reaches a two-cycle.
}
\end{figure}

Again we consider the emerging return map
as done before in Fig.~\ref{fig2} for the special case of
$p=0.5$. Lowering $p$
reduces the number of units stimulus ${\bf 1}$ is mapped to, thereby
increasing the number of units stimulus ${\bf 0}$ is mapped to. Comparing
with Fig.~\ref{fig2}(e) the unfilled branch of the return map becomes
steeper
whereas the filled branch becomes flatter. For values of $p\leq0.1$ typically
a return map as shown in Fig.~\ref{fig4}(a) develops. Here one branch of
the map is a constant (horizontal line). such that after presentation of the
infrequent stimulus the centre of activity $i^\ast(t)$ does not depend on the 
previous one $i^\ast(t-1)$. As illustrated by Fig.~\ref{fig4}(b) the
network state $i^\ast$ passes a transient and reaches an attractor provided
persistent presentation of the frequent stimulus. The network state is a
measure of the time having passed since the last presentation of the
infrequent stimulus.

In summary, we have formulated and examined a simple model of memory dynamics based on
a few asumptions. We have shown that the dynamics based on these assumptions
readily
builds up a structure for systematic storage of recent stimuli. We have
also demonstrated the adaptation of the memory in reaction to the information
contained in the stimuli.
Importantly, no correlations in the stream of stimuli are required for the
structure to emerge. A neural network can learn a basic spatial representation
of temporal information before temporally correlated information itself is available. Noise is enough in order to build up a memory.

\end{document}